\documentclass[12pt]{article}
\usepackage{graphicx}
\usepackage{amsmath}
\usepackage{longtable}

\begin{document}

% 4 rows Astronomy Letters style:
\makeatletter
\renewcommand*{\@cite}[2]{{#2}}
\renewcommand*{\@biblabel}[1]{#1.\hfill}
\makeatother

\title{3D Stellar Reddening Map from 2MASS Photometry: An Improved Version}
\author{George~A.~Gontcharov\thanks{E-mail: george.gontcharov@tdt.edu.vn}}

\maketitle

1. Department for Management of Science and Technology Development, Ton Duc Thang University, Ho Chi Minh City, Vietnam\\
2. Faculty of Applied Sciences, Ton Duc Thang University, Ho Chi Minh City, Vietnam


An improved version of the 3D stellar reddening map in a space with a radius of 1200 pc around the Sun and within 600 pc of the Galactic midplane is presented. 
As in the previous 2010 and 2012 versions of the map, photometry with an accuracy better than $0.05^m$ in the $J$ and $K_s$ bands for more than 70 million
stars from the 2MASS catalogue is used in the new version. However, the data reduction technique is considerably more complicated. 
As before, an analysis of the distribution of stars near the main-sequence turnoff on the $(J-K_s)$ - $K_s$ diagram, where they form a distribution maximum, 
provides a basis for the method. The shift of this maximum, i.e., the $mode(J-K_s)$, along $(J-K_s)$ and $K_s$, given the spatial variations of the mean de-reddened
color $(J-K_s)_0$ of these stars, is interpreted as a growth of the reddening with increasing distance. 
The main distinction of the new method is that instead of the fixed mean absolute magnitude, de-reddened color, distance, and reddening for each cell, the individual 
values of these quantities are calculated for each star by iterations when solving the system of equations relating them. This has allowed one to increase the random 
accuracy of the map to $0.01^m$ and its spatial resolution to 20 pc in coordinates and distance and to $1^{\circ}$ in longitude and latitude. 
Comparison with other reddening estimates for the same spatial cells and Gaia DR1 TGAS stars shows that the constructed map is one of the best maps for the
space under consideration. Its systematic errors have been estimated to be $\sigma(E(J-K_s))=0.025^m$, or $\sigma(E(B-V))=0.04^m$. 
The main purpose of the map is to analyze the characteristics of Galactic structures, clouds, and cloud complexes. 
For this purpose, the reddening map within each spatial cell has also been computed by analyzing the reddening along each line of sight.

\newpage

\section*{Introduction}

The sky surveys with accurate stellar photometry made in recent years are actively used to determine the reddening for millions of stars and the interstellar
extinction toward them and to construct 3D reddening and extinction maps as a function of Galactic coordinates. For example, Berry et al. (2012), Kunder et al. (2017), 
and Green et al. (2015) determined the reddening and extinction based on photometry from the SDSS and 2MASS catalogues (Skrutskie et al. 2006), 
the RAVE5 catalogue, and the Pan-STARRS1 and 2MASS catalogues, respectively.
2MASS is a key catalogue in many such studies, because it contains an important source of information about the spectral energy distribution of a star --
near infrared (IR) photometry, in the $J$, $H$, and $K_s$ bands at wavelengths of 1.25, 1.65, and 2.16 $\mu$m, respectively.

However, many of the present-day sky surveys do not contain accurate photometry for the brightest stars, because these stars are too bright for the recording equipment. 
The overwhelming majority of giants within several hundred parsecs of the Sun also belong to such stars, but a 3D reddening and extinction map for this space is very topical, 
and it can be constructed using the photometry of mainsequence (MS) stars less bright than giants. However, this requires solving another problem: the giants
and MS stars are not separated when using visual photometry because of a significant visual reddening.
Only IR photometry allows them to be separated. The direct method of their separation, through the calculation of absolute magnitudes from trigonometric
parallaxes, will apparently be effective after the appearance of parallaxes for all stars of the Gaia project. However, the data from Gaia DR1 TGAS (Gaia 2016), 
which actually contains the stars from the Tycho-2 catalogue (H\o g et al. 2000), are not enough to construct a 3D reddening and extinction map, 
because the parallaxes in it have a sufficient accuracy and the samples of dwarfs and giants are complete in a very small solar neighborhood. 
For example, Gontcharov (2008b, 2011, 2016a (Table 1)) obtained complete samples of clump and branch giants from Tycho-2 and showed that they could be
complete only to 600 and 760 pc, respectively, toward the Galactic poles and in an even smaller solar neighborhood in other directions. Thus, one of the
best sources of data on the reddening and extinction in Galactic solar neighborhoods is IR photometry for MS stars.

Using 2MASS photometry in the $J$ and $K_s$ bands with an accuracy higher than $0.05^m$ for MS stars, Gontcharov (2010) constructed a 3D reddening
$E(J-K_s)$ map for each cubic $100\times100\times100$ pc spatial cell along the Galactic rectangular coordinates $X$, $Y$, and $Z$ within 1600 pc of the Sun.

For the second version of the same map Gontcharov (2012b) reduced the same 2MASS photometric data by the same method but used a moving
averaging to increase the spatial resolution of themap to $50\times50\times50$ pc. The reddening $E(J-K_s)$ was calculated for the same $100\times100\times100$ pc
cells as those in the first version of the map, but each cell was shifted by 50 pc instead of 100 pc along each of the rectangular coordinates. 
Then, the relation
\begin{equation}
\label{rllaw}
E(B-V)=1.9E(J-K_s)
\end{equation}
in accordance with the extinction law from Rieke and Lebofsky (1985) at $R_V=3.1$, was used to determine $E(B-V)$. The product of the second version of the
3D reddening map by the 3D map of spatial variations in the coefficient $R_V$ from Gontcharov (2012a), in accordance with the relation
\begin{equation}
\label{are}
A_V=R_VE(B-V)
\end{equation}
allowed Gontcharov (2012b) to produce a 3D extinction $A_V$ map with a resolution of 50 pc and an accuracy $\sigma(A_V)=0.2^m$.

In this study we constructed the third, improved version of the map. It is based on the same photometric data but was reduced by a more complex method.

\section*{A general description of the method}

Figure 1a shows the Hertzsprung--Russell (HR) $(J-K_s)$ -- $M_{K_s}$ diagram for 33540 stars from the Gaia DR1 TGAS catalogue with a relative parallax
error $\sigma(\varpi)/\varpi<0.15$ (according to the recommendation of the TGAS authors, the uncertainty of 0.3 milliarcseconds (mas) describing the disregarded
systematic errors of $\varpi$ was added to the formal error $\sigma(\varpi)$ specified by them), 2MASS photometry in the $J$ and $K_s$ bands more accurate than $0.05^m$, 
and the constraint $|b|>65^{\circ}$ to minimize the influence of reddening $E(J-K_s)$ and extinction $A_{K_s}$. The stellar positions in the figure were not corrected for 
the reddening and extinction. This is justifiable, because for so high latitudes this study, in agreement with others, showed $E(J-K_s)<0.06^m$ and $A_{K_s}<0.03^m$, i.e.,
at the level of photometric errors.

For the data in Fig. 1a we use the stellar distances that were calculated from the Gaia DR1 TGAS parallaxes by Astraatmadja and Bailer-Jones (2016). 
They showed that, in general, calculating the distance as $R=1/\varpi$ is incorrect if $\sigma(\varpi)$ are known. However, for the Gaia DR1 TGAS stars with 
$\sigma(\varpi)/\varpi<0.15$ considered here the difference between the distances calculated by the above two methods is insignificant.

\begin{figure}
\includegraphics{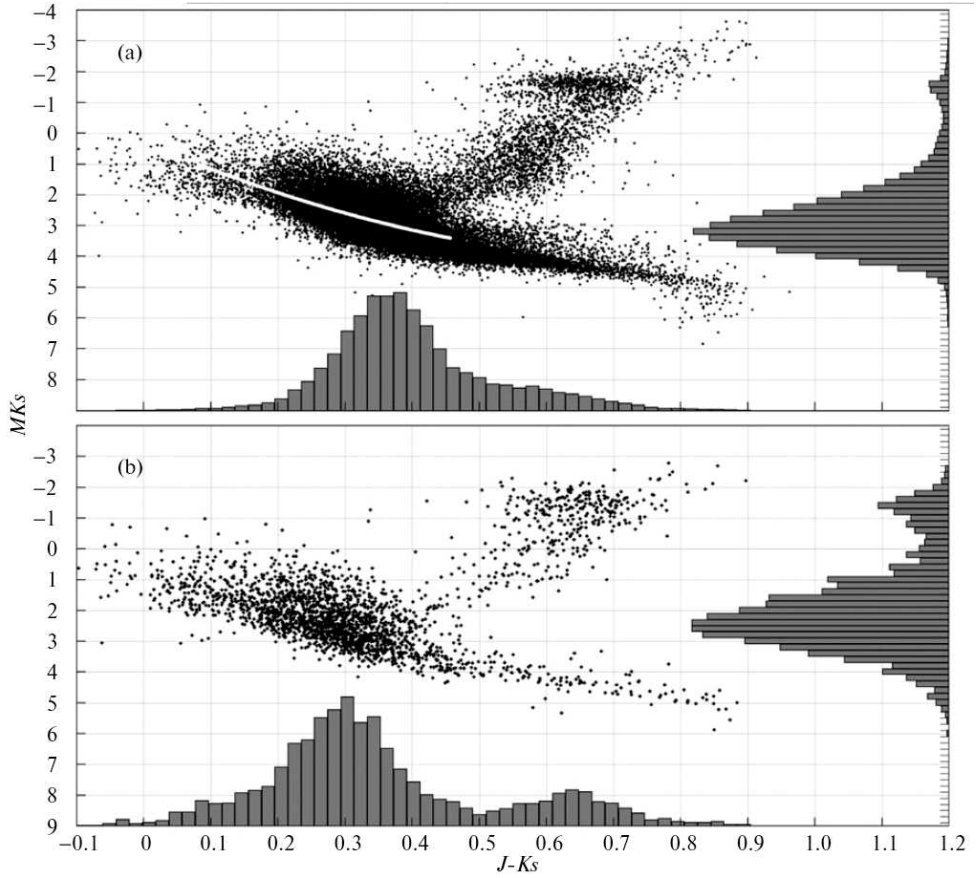}
\caption{Hertzsprung--Russell diagram for (a) 33540 stars from the Gaia DR1 TGAS catalogue with $\sigma(\varpi)/\varpi<0.15$ and $|b|>65^{\circ}$
and (b) 2434 stars from the Hipparcos catalogue with $\sigma(\varpi)/\varpi<0.15$ and $|b|>65^{\circ}$. The values of $\varpi$ from the corresponding catalogue were used. 
The white curve indicates the dependence (9). The histograms of the distribution of stars in $(J-K_s)$ and $M_{K_s}$ are shown on the lower right. 
The stellar positions were not corrected for the reddening $E(J-K_s)$ and extinction $A_{K_s}$.
}
\label{hr}
\end{figure}

Only 5568 of these stars (17\%) are Hipparcos stars (van Leeuwen 2007), while only 2434 stars are selected directly from Hipparcos under the same
constraints. An analogous HR diagram for them is shown in Fig. 1b (the Hipparcos parallaxes are used). The TGAS catalogue not only increased considerably
the number of stars with accurate parallaxes but also shifted the modes $(J-K_s)$ and $M_{K_s}$. This can be seen from a comparison of the histograms of the distribution
of stars in these quantities on the lower right on each panel. The reason is the selection in favor of bright and, consequently, blue stars in Hipparcos, which disappears in TGAS.

In both cases, however, the maximum of the distribution in both quantities falls on the stars located near the MS turnoff to the giant branch, i.e., F--G stars. 
Apart from single dwarfs of approximately solar metallicity and age, this distribution maximum is also formed by binary stars, low-metallicity subdwarfs,
and subgiants. These categories of stars increase noticeably the dispersion of $(J-K_s)$ and $M_{K_s}$ in this part of the diagram. In the range
$0.2^m<J-K_s<0.45^m$ we have $\sigma(M_{K_s})=0.67^m$ against $0.76^m$, respectively, for the stars selected by the TGAS and Hipparcos parallaxes. 
The smaller scatter of TGAS stars is attributable to more accurate $\varpi$. However, according to Parenago (1954), an error $\sigma(\varpi)/\varpi<0.15$ 
causes an error $\sigma(M_{K_s})<0.33^m$. Consequently, not the error in $\varpi$ but the real diversity of stars makes a major contribution to $\sigma(M_{K_s})$.

It can be seen from the histograms that most of the stars near the MS turnoff lie within the ranges $0.28^m<(J-K_s)<0.44^m$, $2.2^m<M_{K_s}+A_{K_s}<4.0^m$
($M_{K_s}+A_{K_s}$ rather than $M_{K_s}$, because the extinction on the diagram is disregarded). Given that the 2MASS photometry in the $J$ and $K_s$ bands has an
accuracy higher than $0.05^m$ almost for all stars with $5^m<K_s<14^m$,we have the constraint $36<R<1000$ pc. The softer constraint $M_{K_s}+A_{K_s}<3.6^m$
gives $R<1200$ pc and a still fairly complete sample of stars. On the other hand, the density of F--G stars far from the Galactic midplane is insufficient to construct
the map by this method, which is also confirmed by theoretical estimates, for example, according to the Besan\c{c}on model of the Galaxy (Czekaj et al. 2014).
Therefore, we finally adopted the space of radius $R=1200$ pc around the Sun with the constraint $|Z|<600$ pc for our consideration in this study.
It was divided into 630109 cubic $20\times20\times20$ pc cells, in each of which the reddening $E(J-K_s)$ was calculated.

In addition, the reddening $E(J-K_s)$ was calculated by the same method and in the same space but as a function of spherical coordinates $R$, $l$, and $b$ for
cells with a depth of 20 pc in $R$ and a width of 1$^{\circ}$ in $l$ and $b$. This presentation of the results allows us to construct not only the cumulative reddening map
from the observer to the cell under consideration but also the differential reddening map in a specific spatial cell from the near to the far boundary of the cell. This
map is constructed and considered below. In addition, this presentation of the results allows the variations in reddening, de-reddened color, and other quantities along
various lines of sight to be considered.

However, the cases where the reddening on one line of sight in some range of distances decreases with distance are encountered. As was pointed out by
Gontcharov (2010) when producing the first version of the map, this is an inevitable consequence of the finite and moderately low spatial resolution of the
map. In this situation, the cases where the spatial cell is slightly larger than the absorbing cloud contained in it are quite frequent. Stars farther than the cloud
but outside its projection onto the celestial sphere are then observed in the part of the cell unoccupied by the cloud. Their reddening is usually smaller than
that for the stars inside the cloud. As a result, a local reddening maximum corresponding to the cloud and then a decrease in reddening with distance will
be seen on such a line of sight. This decrease by no means implies that the reddening immediately behind the cloud is smaller than that inside it. Attempts to
remove this effect by increasing the spatial resolution of the map have not yet yielded a guaranteed result: even the creators of the 3D map with the highest
angular resolution to date, of the order of a few arcminutes (Green et al. 2015), provide no evidence that the reddening variations on one line of sight found by
them are caused by the distance errors rather than by the effect being considered here. The key reason for the difficulties in removing this effect is a very small
scale of the natural fluctuations in the properties of the interstellar medium: these fluctuations, obviously, are significant on a scale smaller than 10 pc and even,
possibly, within each parsec.

As in the previous versions of the map, in this study the shift of the maximum of the distribution of MS turnoff stars along $(J-K_s)$ and $K_s$, i.e., the
modes of these quantities, given the spatial variations of the mean de-reddened color $(J-K_s)_0$ of these stars, is interpreted as a growth of the reddening
\begin{equation}
\label{ejk}
E(J-K_s)=(J-K_s)-(J-K_s)_0
\end{equation}
with increasing distance $R$. In other words, in this method two 3D maps are constructed for the cells of the space under consideration: $(J-K_s)$ and $(J-K_s)_0$. 
In accordance with Eq. (3), their difference is a 3D reddening map.

The proportionality of the means $K_s$ and $R$ for the spatial cells underlies this interpretation. Indeed, for each spatial cell
\begin{equation}
\label{gon1}
\overline{R}=10^{(\overline{K_s}-\overline{M_{K_s}}+5-\overline{A_{K_s}})/5}.
\end{equation}

In the previous versions of the map in the entire space under consideration we assumed the extinction $A_{K_s}$ to be negligible and the mean $M_{K_s}$
for the MS turnoff stars to be constant. Then, Eq. (4) turns into a simple relationship between $\overline{K_s}$ and $\overline{R}$.

In contrast to the previous versions, in this version the interrelated characteristics of each star in the range $0.2^m<(J-K_s)<0.8^m$ are refined by iterations
when solving the system of equations
\begin{equation}
\label{sist}
\left\{
\begin{aligned}
A_V=f_1(R, l, b)\\
A_{K_s}=0.117A_V\\
R=10^{(K_s-A_{K_s}-M_{K_s}+5)/5}\\
(J-K_s)_0=f_2(R, b)\\
M_{K_s}=f_3(J-K_s)_0\,,
\end{aligned}
\right.
\end{equation}
where $f1$, $f2$, and $f3$ are some functions to be considered below. This approach is similar to that applied by Gontcharov (2012c) when investigating a sample of
OB stars.

In the space under consideration both the maximum of the distribution of MS turnoff stars and the part of the ``wings'' of this distribution necessary
for identifying the maximum always fall within the range $0.2^m<(J-K_s)<0.8^m$. Apart from this maximum, the second maximum comprised by clump
giants is also seen in Fig. 1. In the entire space under consideration the color index $(J-K_s)$ allows precisely the sought-for left (the bluest, with smaller $(J-K_s)$) 
maximum of the distribution of stars to be identified quite efficiently. This is illustrated by Fig. 2, where the distribution of 2MASS stars by $0.01^m$-wide
$(J-K_s)$ cells at various distances is shown for one direction $l\approx102^{\circ}, b\approx5^{\circ}$: (a) 205, (b) 307, (c) 613, and (d) 1193 pc. 
The left, central, and right arrows mark, respectively, $(J-K_s)_0$, $mode(J-K_s)$ for the MS turnoff stars, and the limiting $(J-K_s)$
to which $mode(J-K_s)$ is sought. A shift of these quantities with distance is seen. We also see that  $mode(J-K_s)$ is determined reliably, with an accuracy
of $0.01^m$. At $R=205$ and 307 pc we see the second maximum at $(J-K_s)>0.6^m$ produced by clump giants. Therefore, at small $R$ it is particularly
important to preestablish the limiting  $(J-K_s)$ to which $mode(J-K_s)$ is sought. For this purpose, we used an estimate based on the 3D model of spatial $A_V$
variations from Gontcharov (2009, 2012b). The limiting $(J-K_s)$ was established with a large margin so as to reliably identify the dip in the distribution after
the sought-for peak (between the central and right arrows in Fig. 2).

Reddening-shifted OB stars from one of the largest OB associations, Cep~OB2, manifested themselves at $R=613$ pc: in Fig. 2c they are enclosed by
the ring. We see that they cannot affect the position of the distribution maximum, i.e., $mode(J-K_s)$. An excess of OB stars in this direction manifests itself
in the range $613<R<697$ pc. The distance to the association Cep~OB2 found in this way is consistent with the independent estimates by de Zeeuw
et al. (1999) and Gontcharov (2008a), respectively, $615\pm35$ and $766\pm47$  pc. Thus, a byproduct of this method can be the determination of the distances and
reddenings for OB associations and open clusters.

\begin{figure}
\includegraphics{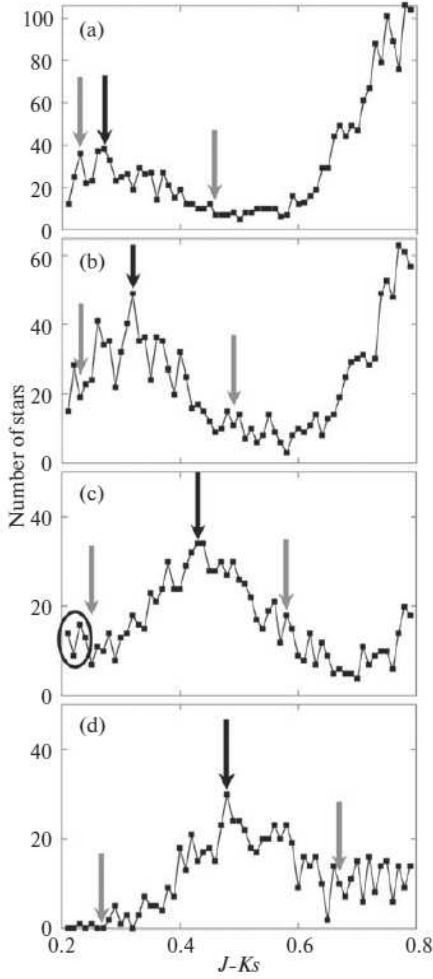}
\caption{Distribution of 2MASS stars by $0.01^m$-wide $(J-K_s)$ cells for the spatial cells toward the OB association Cep~OB2 at a distance of (a) 205, (b) 307, (c) 613, and
(d) 1193 pc. The ring encloses a group of OB stars from this association. The left, central, and right arrows mark, respectively, $(J-K_s)_0$, $mode(J-K_s)$ for the
MS turnoff stars, and the limiting $(J-K_s)$ to which $mode(J-K_s)$ was sought.
}
\label{jkn}
\end{figure}

To construct a map with comparatively smooth reddening variations and a high spatial resolution and to smooth out the influence of the errors in $R$, we took
into account each star not in one but in 19 spatial cells: one with coordinates $X$, $Y$, $Z$ and 18 nearest to it. Thus, a spherical smoothing window with a
radius of $28.3$ pc was actually applied. This does not contradict the uniqueness of the determination of $R$ and $Z$ for each star, because the accuracy of
their determination for one star is comparatively low, poorer than 30 pc almost everywhere. We applied an analogous method when producing the $E(J-K_s)$
map as a function of $R$, $l$, and $b$. However, since here the linear sizes of the cell change significantly at its fixed angular size, the number of cells involved in
the averaging was changed so that at least five stars fell into the cell of 4--dimensional space $R-l-b-mode(J-K_s)$.

When solving the system of equations (5), in each step of iterations we sorted again the stars by spatial cells and within each of them by 60 $(J-K_s)$ cells
in the range $0.2^m<(J-K_s)<0.8^m$ with a $0.01^m$ step. Therefore, naturally, in each step of iterations the maximum of the distribution of stars and the
sought-for $mode(J-K_s)$ can shift both in the space and along $(J-K_s)$, because $R$ and $(J-K_s)_0$, followed by $X$, $Y$, $Z$, and $(J-K_s)$, can change. 
In all cells of the space under consideration the solution of the system of equations (5) converges to an unambiguous result after no more than 13 iterations.
As in the previous versions of the map, the reddening $E(J-K_s)$ in a spatial cell is calculated from Eq. (3) for $mode(J-K_s)$.

The main advantage of the method under consideration is that no stars are selected for the analysis and, therefore, the corresponding random and systematic
errors do not emerge.

In the system of equations (5) the relation
\begin{equation}
\label{akav}
A_{K_s}=0.117A_V
\end{equation}
along with the relations
\begin{equation}
\label{parseclaw}
E(B-V)=1.655E(J-K_s),
\end{equation}
\begin{equation}
\label{av31ebv}
A_V=3.1E(B-V),
\end{equation}
in contrast to the previous versions of the map with relation (1), were taken from the extinction law adopted in the PARSEC database of theoretical
isochrones (Bressan et al. 2012). In the wavelength range $0.4<\lambda<2.2$ $\mu$m between the $B$ and $K_s$ bands important for us this extinction law slightly
differs from other popular laws, for example, from Cardelli et al. (1989), even at fixed $R_V=3.1$. The spatial variations of the extinction law and the large
width of the bands under consideration introduce an additional uncertainty into the extinction law for a specific star or spatial cell. Therefore, it worth
emphasizing that the reddening $E(J-K_s)$ estimates are a direct result of this study and the main content of the presented map. We recommend to use them
in all possible cases, while the reddening $E(B-V)$ and extinction $A_V$ estimates based on them are secondary.

When solving the system of equations (5) the function $f1$ was taken from the very beginning as the previously mentioned 3D analytical model of spatial
$A_V$ variations as a function of $R$, $l$, and $b$ from Gontcharov (2009, 2012b). However, as a result of solving the system of equations (5), the function $f1$
deviated noticeably from the initial values and actually, if the fixed relations (7) and (8) are taken into account, is the sought-for result, the dependence of
reddening on coordinates in tabular form.

In the system of equations (5) the function $f3$ describes the MS in its middle part. The empirical relation adopted below as the function $f3$,
\begin{equation}
\label{jk0mks}
M_{K_s}=0.22+9.43(J-K_s)_0-5.57(J-K_s)_0^2
\end{equation}
was calculated by the least-squares method for TGAS stars in the range $0.1^m(J-K_s)<0.45^m$ and is indicated by the white curve in Fig. 1. 
The coefficients of Eq. (9) were calculated for different cases: with and without giants, with various color constraints for the sample of stars, and with the
polynomial order increased to 4. The difference between the 3D maps constructed in these cases for most spatial cells is small: $\Delta E(J-K_s)<0.01^m$.
Close coefficients of Eq. (9) were also found in the simulations of stellar characteristics considered below. In addition, the dependence (9) was also
calculated for stars with accurate $\varpi$ from Hipparcos. It is shifted by $\Delta M_{K_s}=0.12^m$ in the sense of ``TGAS minus Hipparcos'', i.e., 
according to the TGAS data, the stars have a lower luminosity. This is caused by the selection in the Hipparcos catalogue in favor of bright stars.

\section*{The spatial $(J-K_s)_0$ variations}

The function $f2$ in the system of equations (5) reflects the spatial variations in the de-reddened color of MS turnoff stars and is actually the previously mentioned
3D map of these variations in analytical form being subtracted from the 3D map of $(J-K_s)$ variations in accordance with Eq. (3).

The search for the function $f2$ and the entire method under consideration are based on the assumption that the mean de-reddened color $(J-K_s)_0$
of the stars constituting the distribution maximum on the $(J-K_s)$ -- $K_s$ or $(J-K_s)$ -- $M_{K_s}$ diagram changes smoothly and predictably in the space
under consideration. This, in turn, is based on the assumption that (1) the corresponding stellar population in the space under consideration changes
its mean properties smoothly, predictably, and by a small amount and (2) the 2MASS catalogue is fairly complete in the space under consideration
up to the limiting magnitude under consideration ($K_s<14^m$). The latter assumption is important, because the de-reddened color $(J-K_s)_0$ must change
significantly with $R$ due to the selection of stars. These assumptions limit the region of space where the method under consideration is applicable.

\begin{figure}
\includegraphics{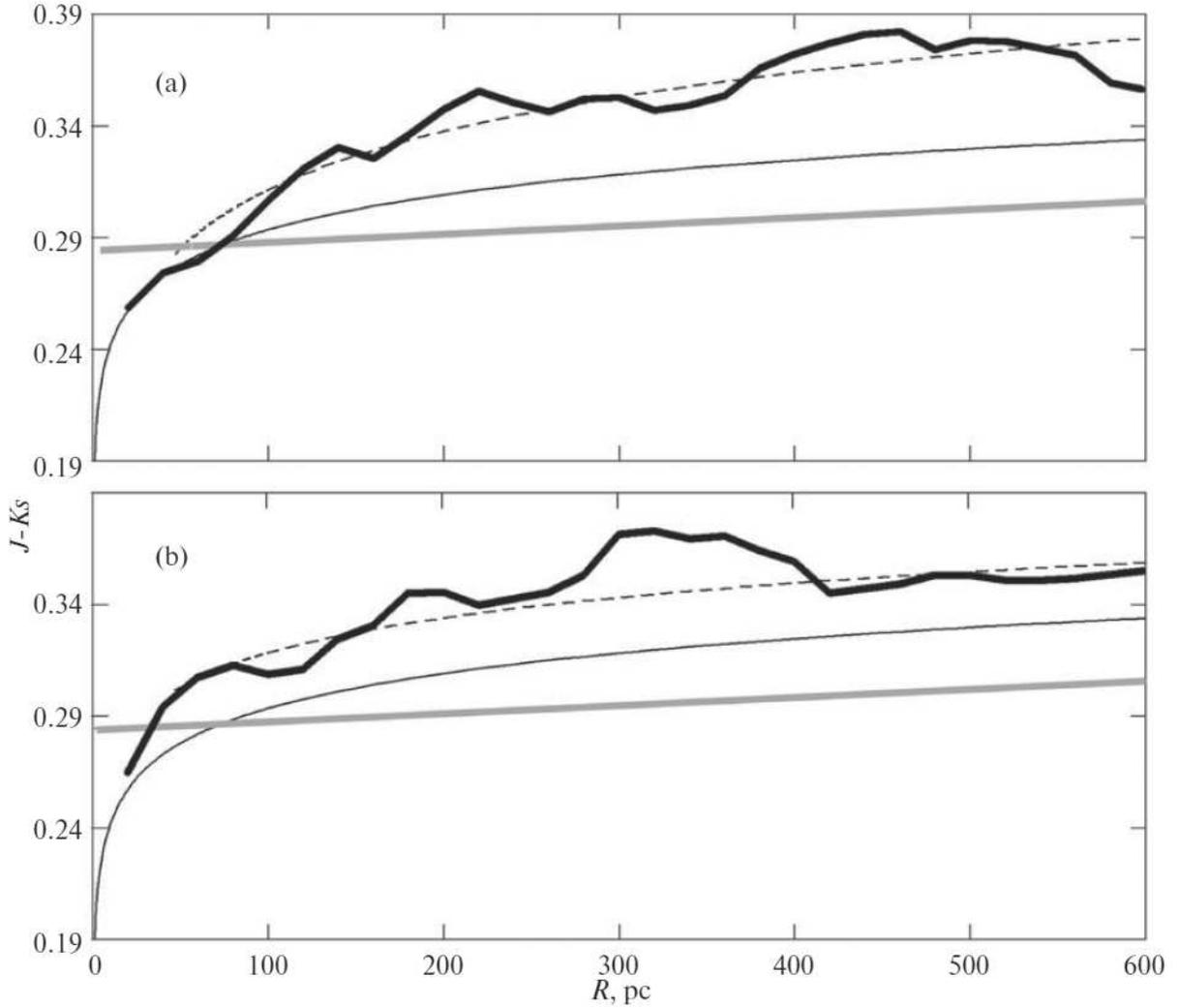}
\caption{$(J-K_s)$ versus $R$ toward the north (a) and south (b) Galactic poles (black thick polygonal curves); the fit to this
dependence by the logarithm of $R$ (black dashed curves); the adopted logarithmic dependence (11) (black smooth curves); the
dependence (10) in the previous version of the map (with a different extinction law) (gray thick straight lines).
}
\label{ngpsgp}
\end{figure}

Gontcharov (2010) tested these assumptions for real Hipparcos stars and in Monte Carlo simulations of the characteristics of2MASS stars located near the
MS turnoff in a space with a radius of 1.6 kpc around the Sun and having $K_s<14^m$. These simulations are based on the first version of the the Besan\c{c}on
model of the Galaxy (Robin et al. 2003) and the theoretical PARSEC data (Marigo et al. 2008). By now, both the Besan\c{c}on model (Czekaj et al. 2014)
and the PARSEC data (Bressan et al. 2012) have been revised considerably. Therefore, for the new version of the map the simulations were performed
again. The dependences of all the quantities under consideration and the random number generator remained as before. As has been noted previously, the
new simulations confirmed the dependence (9). In addition, the new simulations gave a slightly different representation of the function $f2$ than that in the
previous versions of the map.

In the previous versions of the map, as a result of the simulations and comparison with real data, we adopted the spatial variations of the de-reddened color as
a function of b and R in pc in accordance with the equation
\begin{equation}
\label{gon3}
(J-K_s)_0=0.23+0.09\,\sin|b|+0.00004\,R\,.
\end{equation}
The coefficients in this equation were chosen so as to minimize the reddening for all lines of sight provided that the reddening:
\begin{enumerate}
\item smoothly approaches 0 when R approaches 0;
\item is negative only in a statistically insignificant number of spatial cells;
\item decreases with R only locally, i.e., $E(J-K_s)$ on each line of sight can be represented as a nondecreasing function of $R$.
\end{enumerate}

Given the fairly high accuracy of the photometry (in the entire space the median of the error in each band is $0.02^m$) and the comparatively large number of
stars in a typical cell (on average, 21 stars fell into a cell of 4--dimensional space $X-Y-Z-mode(J-K_s)$, see also Fig. 2), the second condition implies
that a negative reddening on each line of sight is admissible only in the region $|R|<40$ pc, where, as has been noted previously, the 2MASS photometry is inaccurate.

The new simulations led us to conclude that the de-reddened color $(J-K_s)_0$ should be proportional to $\log(R)$ rather than $R$. Thus, the nondecreasing function
of $R$ in the third condition is a logarithm. In this study we obtained the following as the function $f2$ instead of Eq. (10):
\begin{equation}
\label{gon4}
(J-K_s)_0=0.10+0.09\,\sin|b|+0.0518\,\log(R).
\end{equation}

Just as for the previous versions of the map, the coefficients in Eq. (11) were chosen by the leastsquares method so as to minimize the reddening
for all lines of sight under the previously specified conditions. In fact, the most stringent constraints on the sought-for coefficients are imposed by the
data used on the lines of sight with minimum reddening, i.e., in the known sky regions marked by Schlegel et al. (1998) in their Table 5. These regions
also include the Galactic poles, although Schlegel et al. (1998) showed that the reddening is minimal not in them, while Gontcharov (2016b) showed that
the regions of minimum reddening are shifted from the Galactic poles toward the poles of the Gould Belt, which is a dust container.

As an example, the black thick polygonal curves in Fig. 3 indicate the $(J-K_s)$ variations as a function of $R$ toward the north (a) and south (b) Galactic
poles; the black dashed curves indicate the fits to these variations $(J-K_s)=0.137+0.0873\log(R)$ and $(J-K_s)=0.215+0.0518\log(R)$ for the north and
south poles, respectively ($R$ in pc); the black smooth curves indicate the dependence (11); the gray thick straight lines indicate the dependence (10) in the
previous versions of the map with allowance made for the change of the extinction law from (1) to (7). In the large range $R>200$ pc the black smooth curve
and the gray straight line are seen to be approximately parallel to each other. Their vertical shift is mainly caused precisely by the change of the extinction law.
Thus, replacing the dependence of $(J-K_s)_0$ on $R$ in Eq. (10) by the dependence on $\log(R)$ in Eq. (11) had an effect only near the Sun.

At $R>75$ pc the new version of the map for all directions is seen to give larger $(J-K_s)_0$ and, hence, smaller $E(J-K_s)$ that do the previous versions
(i.e., the smooth black curve is higher than the gray straight line). This is because relations (7) and (11) are used instead of (1) and (10).

\begin{figure}
\includegraphics{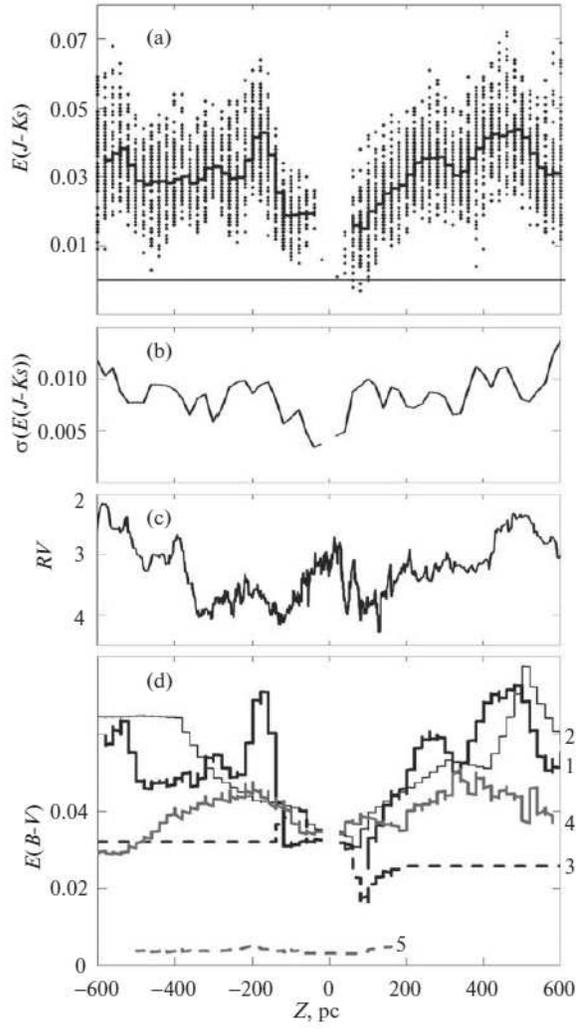}
\caption{Various characteristics versus $Z$ for the spatial cells inside a cylinder with a radius of 100 pc toward the Galactic poles
with the constraint $|b|>45^{\circ}$ (for explanations, see the text).
}
\label{poles}
\end{figure}

The direction toward the south Galactic pole turned out to be the line of sight among all lines of sight on the celestial sphere where a minimum
coefficient at $\log(R)$ was found. Therefore, Fig. 3b clearly illustrates the uniqueness of the determination of dependence (11). We select the slope and vertical
shift of the smooth black curve so as to fulfill the previously mentioned conditions. We see that a smaller slope of the smooth curve would give an
unjustifiable systematic increase in $E(J-K_s)$ (i.e., the difference between the dashed and smooth black curves) at large $R$; conversely, a larger slope of
the smooth curve would give a systematic decrease in $E(J-K_s)$ at large $R$; an upward shift of the smooth black curve would give negative $E(J-K_s)$
(the difference between the black solid curves) at small $R$, while its downward shift would give clearly overestimated values of $E(J-K_s)$ near the Sun
significantly different from zero. Of course, the direction toward the south pole is only one of the lines of sight, where the data imposed significant
constraints on the coefficients of Eq. (11).

Once the lines of sight with minimum reddening have been analyzed, the free term and the coefficient at $\log(R)$ in Eq. (11) were found and fixed in the
subsequent search for the coefficient at $\sin|b|$ by analyzing the data at all latitudes. The accuracy of the coefficients in Eq (11) corresponds to unity of their last digit.

\section*{Errors of the map}

Since very stringent constraints on the sought-for coefficients of Eq. (11) were imposed by the data used in the polar caps, it is interesting to estimate the
final result precisely here. The dots in Fig. 4a indicate the reddening $E(J-K_s)$ obtained in this study as a function of $Z$ for the spatial cells inside a cylinder with
a radius of 100 pc toward the Galactic poles with the constraint $|b|>45^{\circ}$. We see a small number of cells with $E(J-K_s)<0^m$ at $|Z|\le80$ pc 
$(E(J-K_s)=0$ is marked by the horizontal line). The result of our linear filtering by 69 cells is also indicated here by the polygonal curve.

The scatter of points at one $Z$ is caused both by the natural reddening fluctuations between adjacent cells due to the properties of the medium and by the measurement
errors. The standard deviation $\sigma(E(J-K_s))$ as a function of $Z$ is shown in Fig. 4b. We see that $\sigma(E(J-K_s))<0.01^m$ almost everywhere. 
This may be considered as an upper limit for the random errors of the constructed map. However, the systematic errors can be much larger. They include the
errors of the coefficients in Eqs. (11) and (9), which can be estimated in total as $\sigma(E(J-K_s))=0.02^m$ based on the simulation results and the accuracy of
the original data.

In addition, the systematic errors include the error of the adopted extinction law, i.e., the errors of the coefficients in Eqs. (6), (7), and (8). They are difficult
to estimate, because the spatial variations of the extinction law in the space under consideration are known poorly. However, the $E(J-K_s)$ and $E(B-V)$ 
variations found here correlate with the variations of the extinction law. Figure 4c shows the dependence of $R_V$ on $Z$ from the data by Gontcharov (2016) for
branch giants from Tycho-2 in a spatial cylinder along the $Z$ axis with a radius of 150 pc (the curve from his Fig. 1 was reproduced). The peaks of this curve
at $R\approx-570$, $-200$, $+300$ and $+500$ pc are seen to correspond to the maxima of the curve in Fig. 4a, but even better, the black thick curve for $E(B-V)$
as a function of $Z$ in Fig. 4d. Here, for comparison with the result of this study (black thick curve 1), we show the analogous results of filtering the reddening
$E(B-V)$ by 69 cells based on the model by Gontcharov (2012b) (black thin solid curve 2), the model by Arenou et al. (1992) (black dashed curve 3), 
the estimates by Berry et al. (2012) (gray thick solid curve 4), and the estimates by Kunder et al. (2017) (gray dashed curve 5). Here and below, the values of
$E(B-V)$ used for our comparisons were calculated for the spatial cells with a size of 20 pc under consideration using the PARSEC extinction law either
directly (Arenou et al. 1992; Gontcharov 2012b; Kunder et al. 2017) or by interpolating the original data (the remaining sources).

\begin{figure}
\includegraphics{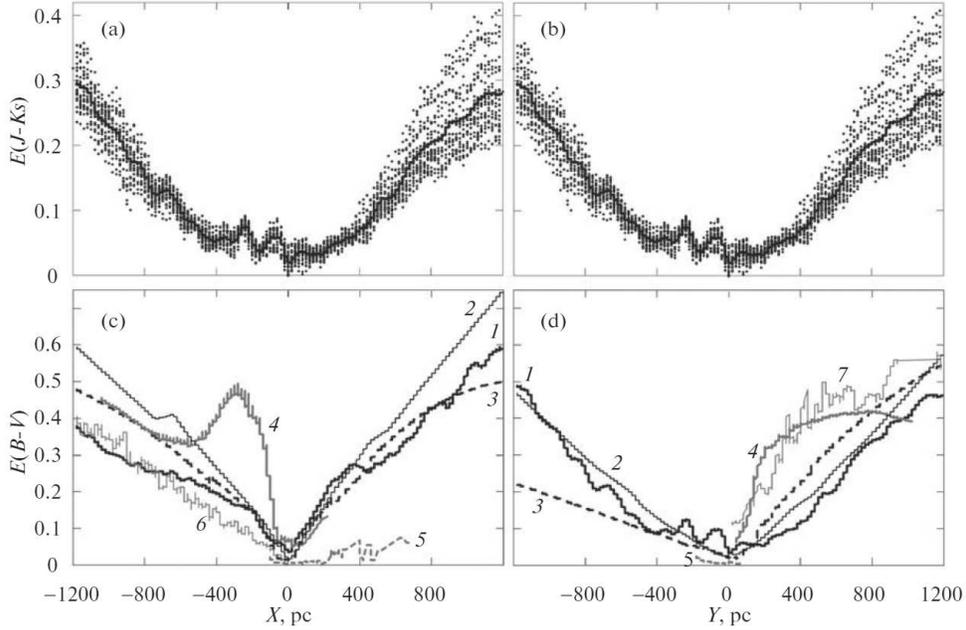}
\caption{$E(J-K_s)$ versus $X$ (a) and $Y$ (b) for the cells inside a cylinder with a radius of 60 pc along the corresponding axis
based on the data of our map (dots), the same data after our linear filtering by 25 cells (black polygonal curve). On panels (c)
and (d) the result of filtering $E(B-V)$ by 25 cells for our map (black thick solid curve 1) is compared with the analogous
results based on the model by Gontcharov (2012b) (black thin solid curve 2), the model by Arenou et al. (1992) (black dashed
curve 3), the estimates by Berry et al. (2012) (gray thick solid curve 4), the estimates by Kunder et al. (2017) (gray dashed curve 5),
the map by Chen et al. (2014) (gray thin solid curve 6), and the map by Sale et al. (2014) (gray thin solid curve 7).
}
\label{ebvxyz}
\end{figure}

The discrepancies in $E(B-V)$ estimates for the polar caps are seen to be small in absolute value: all curves fit into the range $E(B-V)=0.04^m\pm0.04^m$,
which is comparable to the expected errors. The cases of a decrease in $E(B-V)$ with distance seen here are explained by the same errors and variations of the extinction
law. However, because of the small reddening, its relative error reaches 100\%. A large number of extragalactic objects are observed at these latitudes,
and the uncertainty in their characteristics can increase due to the uncertainty in reddening/extinction inside the Galaxy. Consequently, studies are needed
to determine the true reddening and extinction at high latitudes.

The dots in Fig. 5 indicate $E(J-K_s)$ as a function of $X$ (a) and $Y$ (b) for the cells inside a cylinder with a radius of 60 pc along the corresponding
axis based on the data of our map. The same results after their linear filtering by 25 cells are indicated by the black polygonal curves. We see that,
in contrast to the polar caps, $E(J-K_s)$ definitely increases here with heliocentric distance: the local dips do not exceed $E(J-K_s)<0.02^m$. In Figs. 5c
and 5d the result of filtering $E(B-V)$ for our map by 25 cells (black thick solid curve 1) is compared with the analogous results based on the model by
Gontcharov (2012b) (black thin solid curve 2), the model by Arenou et al. (1992) (black dashed curve 3), the estimates by Berry et al. (2012) (gray thick solid curve
4), the estimates by Kunder et al. (2017) (gray dashed curve 5), the map by Chen et al. (2014) (gray thin solid curve 6), and the map by Sale et al. (2014) (gray
thin solid curve 7) (some maps refer only to part of the sky and, therefore, are not presented on all graphs of Figs. 4 and 5). The result by Kunder et al. (2017)
is seen to differ from the remaining ones. It is also obvious that the model by Gontcharov (2012b) overestimates the reddening. This is expectable, because
this model is actually based on the second (2012) version of the map being improved in this study. Its shortcomings, including the use of the extinction
law (1), were discussed previously. In addition, the $E(B-V)$ estimates by Berry et al. (2012) at $R<400$ pc are seen to be far from the remaining estimates
and from reality. The authors themselves concede this. The main reason is that the SDSS photometry is inaccurate for bright (i.e., nearby) stars. Otherwise,
the $E(B-V)$ estimates in Fig. 5 agree well with one another.

The final accuracy of the constructed map can be estimated with some caution as $\sigma(E(J-K_s))=0.025^m$, or $\sigma(E(B-V))=0.04^m$.

The accurate parallaxes of TGAS stars allow the various reddening estimates to be compared by yet another method. For TGAS O--F MS stars with
$\sigma(\varpi)/\varpi<0.1$ (i.e., $R<270$ pc) and accurate (better than $0.05^m$) photometry in $K_s$ and Gaia $G$ bands Fig. 6 shows the de-reddened
color $(G-K_s)_0$ and absolute magnitude $M_G$: (a) uncorrected for $E(G-K_s)$ and $A_G$ and corrected for them based on the PARSEC extinction law 
and the data from (b) this study, (c) the model by Gontcharov (2012b), (d) the model by Arenou et al. (1992), (e) the maps by Chen et al. (2014) and (f) Kunder et al. (2017). 
The results by Berry et al. (2012) and Sale et al. (2014) are not shown, because the former are erroneous at $R<270$ pc, as has been noted previously, while the
latter require a special calibration for stars of early spectral types. The same figure shows the PARSEC isochrones: for an age of 2500 Myr and metallicity
\footnote{To avoid confusion, the metallicity is everywhere denoted by $\mathbf Z$, while one of the Galactic coordinates is denoted by $Z$.}
$\mathbf Z=0.003$ (gray dashed curve), 200 Myr and $\mathbf Z=0.0152$ (gray solid curve), 100 Myr and $\mathbf Z=0.0152$ (black dashed curve), 
10 Myr and $\mathbf Z=0.0152$ (black solid curve). The cross indicates the influence of the photometric and parallax errors. The metallicity $\mathbf Z=0.0152$
is solar (Bressan et al. 2012), while $\mathbf Z=0.003$ is typical for the thick disk. The 100-Myr isochrone may be deemed the zero-age MS (ZAMS). 
The 10-Myr isochrone is shown only partly, leftward of the ZAMS, because it corresponds to stars that have not yet settled on the ZAMS, and the number of such
stars is negligible.

\begin{figure}
\includegraphics{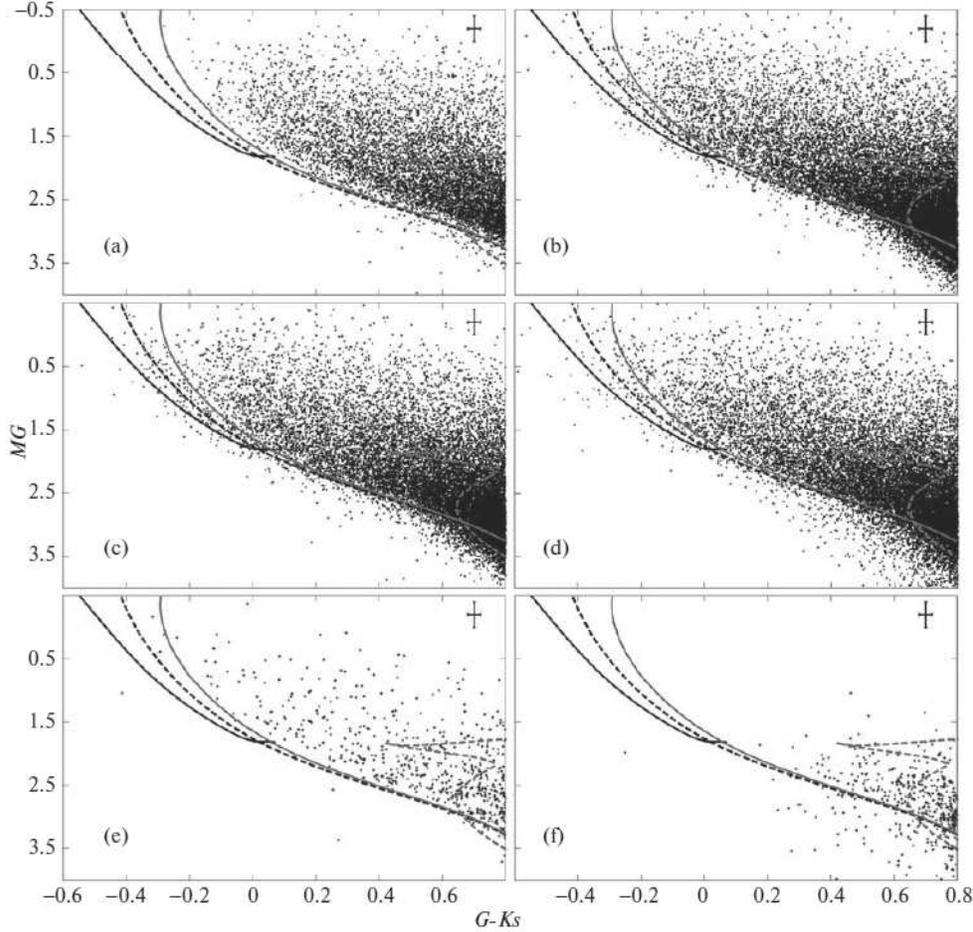}
\caption{Hertzsprung--Russell ``$(G-K_s)_0 - M_G$' diagram for TGAS O--F MS stars with $\sigma(\varpi)/\varpi<0.1$
(à) were not corrected for $E(G-K_s)$ and $A_G$ or were corrected using the PARSEC extinction law and the data on $E(G-K_s)$
and $A_G$ from (b) this study, (c) the model by Gontcharov (2012b), (d) the model by Arenou et al. (1992), (e) the maps by Chen
et al. (2014) and (f) Kunder et al. (2017). The PARSEC isochrones are shown: 2500 Myr, $\mathbf Z=0.003$ (gray dashed curve);
200 Myr, $\mathbf Z=0.0152$ (gray solid curve); 100 Myr, $\mathbf Z=0.0152$ (black dashed curve); 10 Myr, $\mathbf Z=0.0152$ (black solid curve).
The cross indicates the influence of the photometric and parallax errors.
}
\label{gks}
\end{figure}

The number of stars along the 2500-Myr, $\mathbf Z=0.003$ isochrone is large on some graphs. According to Gontcharov et al. (2011), so young low-metallicity subdwarfs
could be formed in the satellites swallowed by the Galaxy. It is also possible that these are ordinary dwarfs with overestimated reddenings.

The HR diagrams based on 2MASS $J$ and $H$, WISE $W2$, and Tycho-2 $B_T$ and $V_T$ photometry appear similar. Too few stars are seen near the 10-,
100-, and 200-Myr isochrones in Fig. 6a (i.e., when the reddening and extinction are ignored). This does not correspond to any model of the Galaxy. The same
is seen in Fig. 6e for the map by Chen et al. (2014). This discrepancy is successfully removed by the map constructed in this study and the models by
Gontcharov (2012b) and Arenou et al. (1992), as can be seen from Figs. 6b, 6c, and 6d, respectively. However, the model by Arenou et al. (1992) gives an
excessively large scatter of stars leftward and below the ZAMS. The reddening estimates in the RAVE project are not too accurate, judging by the large
scatter of points in Fig. 6f. However, only these estimates among all those considered are the direct reddening measurements for the TGAS stars considered
here, while the remaining ones are basically the reddening measurements for neighboring stars interpolated for TGAS stars. Consequently, at the current
accuracy the direct reddening measurements based on an analysis of the spectral energy distribution for a star have no advantages over the indirect estimates
obtained from the reddening of neighboring stars and presented in the form of a 3D reddening map. This emphasizes the topicality of our study.

\section*{Using the map}

To illustrate the results of this study, Fig. 7 shows the the cumulative reddening $E(J-K_s)$ maps as a function of coordinates $X$ and $Y$ to the following
layers: (a) $Z=+90\div+110$, (b) $Z=-10\div+10$, and (c) $Z=-110\div-90$ pc. The Sun is at the center of the graphs. The Galactic center is on the right.
We see the structure of the maps (squares with a 20-pc side) that arose from the division of the space into cells. A similar structure also arises when the space
is divided into cells in $R$, $l$, and $b$. This structure disappears on the contour maps constructed from the same results. They are shown for the same layers
in Figs. 7d, 7e, and 7f, respectively. The correspondence between toning and $E(J-K_s)$ is shown on the right. We see that the small-scale structure
consisting of 20-pc squares does not affect the manifestation of the larger-scale one, the ``rays'' diverging radially from the Sun. This is a natural structure of
any cumulative reddening or extinction map, because each dust cloud on the line of sight causes the reddening/extinction behind it on the same line of sight
to be no smaller than that in it.

\begin{figure}
\includegraphics{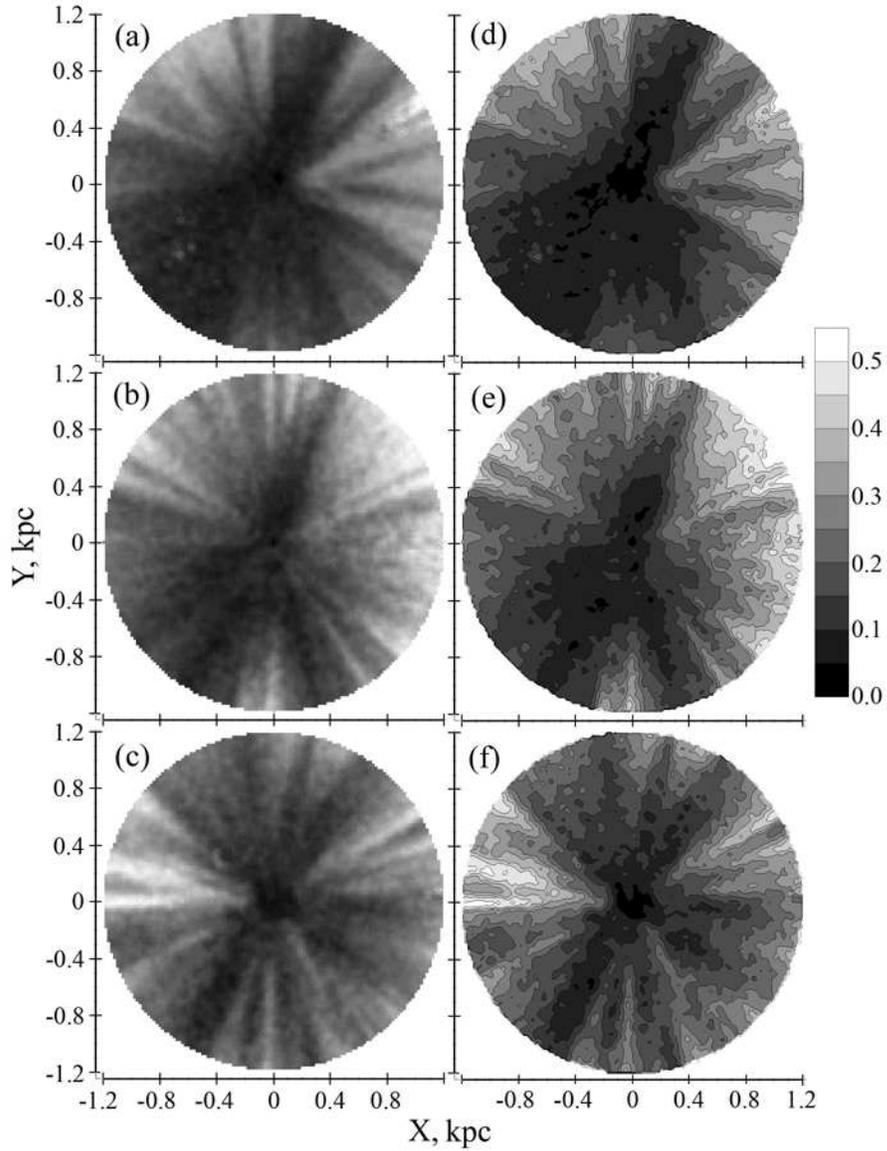}
\caption{Cumulative reddening $E(J-K_s)$ maps as a function of coordinates $X$ and $Y$ to the following layers: (a) $Z=+90\div+110$, (b) $Z=-10\div+10$, and (c) $Z=-110\div-90$ pc.
The contour maps for the same layers (d), (e), and (f), respectively. The Sun is at the center of the graphs. The Galactic center is on the right.
}
\label{xy}
\end{figure}

Comparison of Fig. 7 with the analogous Fig. 3 for the first version of the map (Gontcharov 2010) shows that the new version reveals the same clouds and their
complexes as those in the first version (where their list is given). However, the new version, owing to its higher spatial resolution, gives a much more detailed
picture. The complication of the method also gives hope that this picture is more plausible. In particular, the new version shows much more clearly that above
the Galactic plane (panels (a) and (d)) the vast regions of maximum reddening are located toward the Galactic center and in the second quadrant, while the
regions of maximum reddening toward the Galactic anticenter gradually manifest themselves when passing to the regions below the Galactic plane (panels (c)
and (f)). This is how the well-known structure of the dust distribution in the solar neighborhood manifests itself: apart from the equatorial dust layer, there is
a dust layer in the Gould Belt inclined to the equator approximately by 19$^{\circ}$ that maximally rises above the Galactic plane approximately toward the Galactic
center and sinks toward the Galactic anticenter.

\begin{figure}
\includegraphics{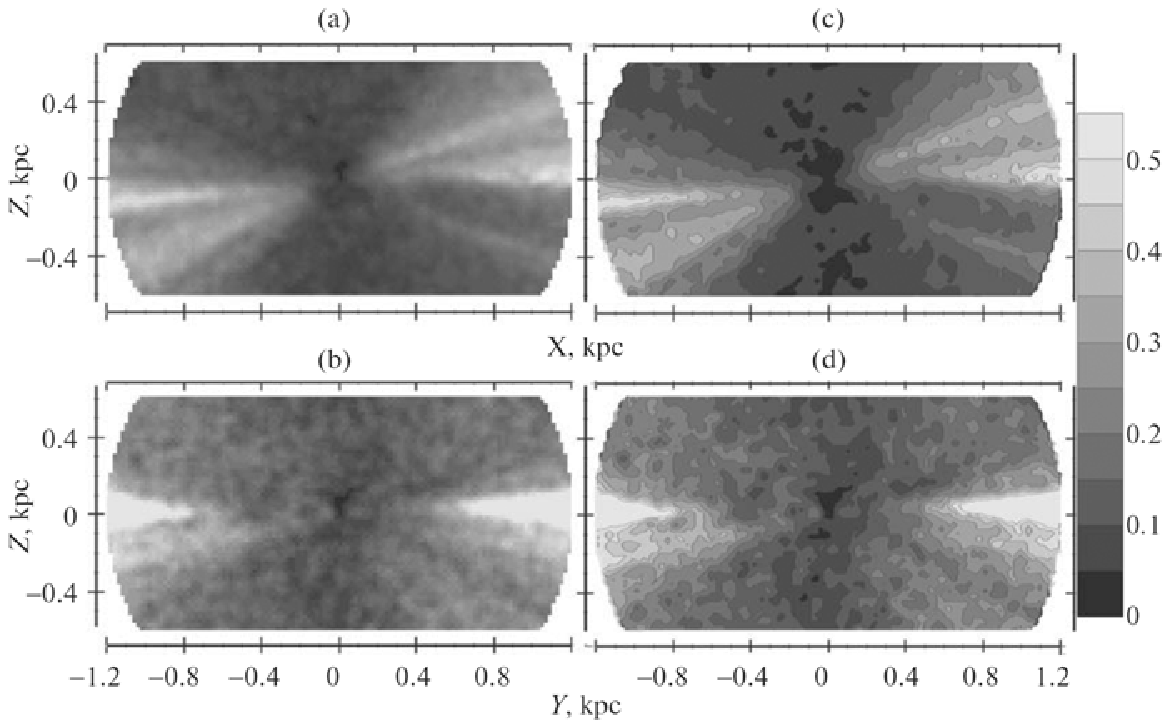}
\caption{Cumulative reddening $E(J-K_s)$ maps (a) as a function of coordinates $X$ and $Z$ to the layer $Y=-10\div+10$ and (b)
as a function of coordinates $Y$ and $Z$ to the layer $X=-10\div+10$. The same on the contour maps (c) and (d), respectively.
The Sun is at the center of the graphs.
}
\label{xyz}
\end{figure}

The dust layer of the Gould Belt is particularly noticeable on the cut of the constructed cumulative reddening $E(J-K_s)$ map in the $XZ$ plane to the
layer $Y=-10\div+10$. This cut is shown in Figs. 8a and 8c as the original and contour maps. The Sun is at the center of the graphs. The Galactic center is
on the right. Four light radial structures are most noticeable here: two approximately toward the Galactic center and two toward the Galactic anticenter. The
pair of structures nearest to the Galactic midplane reflects the equatorial dust layer; the other pair reflects the dust layer of the Gould Belt. In addition, the deviation
of the equatorial dust layer toward the anticenter apparently reflects a general warp of the dust layer in Galactic solar neighborhoods. Figures 8b and 8d
show the original and contour cumulative reddening $E(J-K_s)$ maps in the $YZ$ plane to the layer $X=-10\div+10$. The Sun is at the center of the graphs.

The cumulative reddening map constructed in this study also allows the differential reddening (reddening gradient) map within any small region of space to
be constructed. As an example, Fig. 9 shows the contour reddening $E(J-K_s)$ maps within the layers (a) $R<100$ and (b) $100<R<200$ pc (i.e., for
clarity, five layers with a thickness of 20 pc each were combined on each graph) as a function of coordinates $l$ and $b$. The Galactic anticenter is at the center of
the graph. The toning indicates the gradations with a step $\Delta E(J-K_s)=0.05^m$. Within each toning the isolines indicate the gradations with a step $\Delta E(J-K_s)=0.01^m$. 
In both cases, the maximum reddening gradient is seen to reach $E(J-K_s)=0.11^m$ in the regions near $l\approx10^{\circ}\div25^{\circ}$ marked by the lightest toning.

The darkest toning in Fig. 9b indicates the regions with $-0.03^m<E(J-K_s)<0^m$. Such a value is quite admissible if the above estimate of the accuracy
for the constructed map, $\sigma(E(J-K_s))=0.025^m$, is taken into account. As has been noted previously, such local regions of a decrease in $E(J-K_s)$ with
$R$ appear behind compact dense dust clouds. In Fig. 9a in the layer $R<100$, through which we see the layer $100<R<200$, we see the reddening peaks
(the lightest toning) precisely in the regions of negative $E(J-K_s)$ in Fig. 9b.

\begin{figure}
\includegraphics{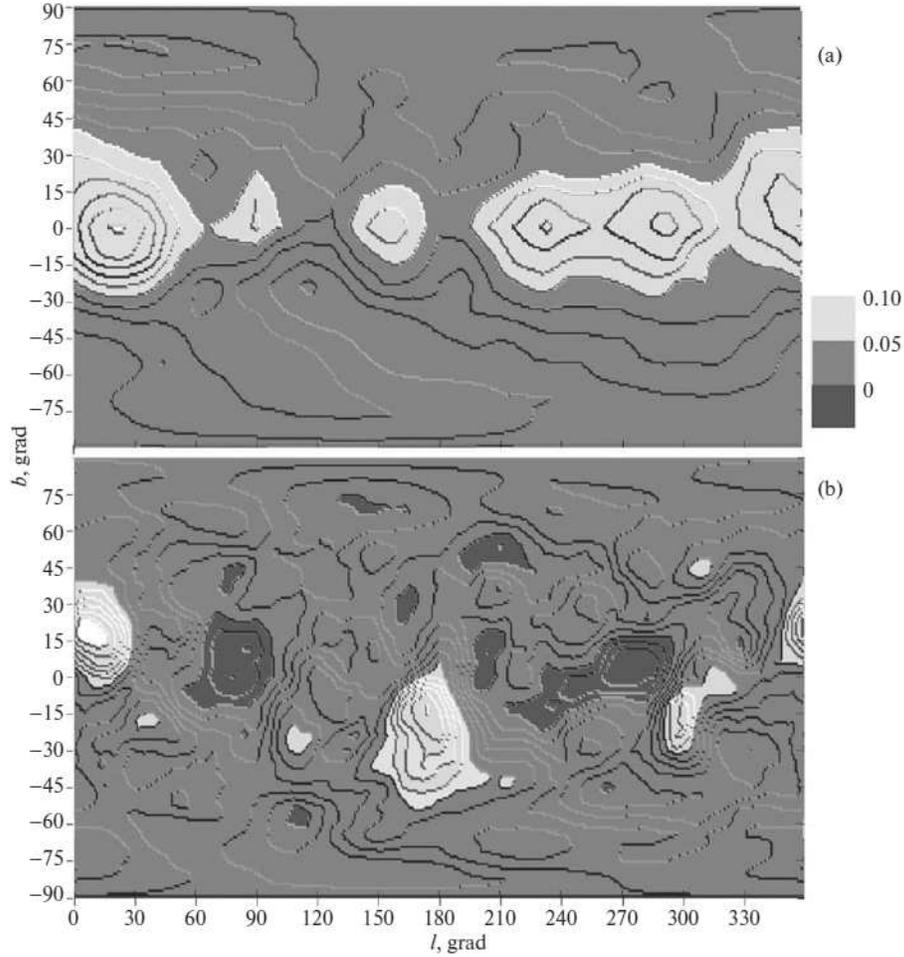}
\caption{Contour differential reddening $E(J-K_s)$ maps within the layers (a) $R<100$ and (b) $100<R<200$ pc as a
function of coordinates $l$ and $b$. The toning indicates the gradations with a step $\Delta E(J-K_s)=0.05^m$. Within each toning
the isolines indicate the gradations with a step $\Delta E(J-K_s)=0.01^m$.
}
\label{dejk}
\end{figure}

The regions with $E(J-K_s)>0.05^m$ (the lightest toning) are located along the equator at $R<100$ pc and deviate noticeably from it at $100<R<200$ pc, 
following the previously discussed orientation of the Gould Belt. At $R<100$ pc the Gould Belt does not manifest itself, because it has a central region
of reduced density of the interstellar medium known as the Local Bubble (Perryman 2009, pp. 324--328, 464--469). In fact, the biggest light spots in Fig. 9b
indicate the cloud complexes of the Gould Belt in this spatial layer: in Ophiucus and Scorpius around $\rho$ Oph ($l\approx10^{\circ}$, $b\approx+20^{\circ}$), 
in Perseus--Taurus--Orion ($l\approx180^{\circ}$, $b\approx-20^{\circ}$), and in Chameleon ($l\approx300^{\circ}$, $b\approx-15^{\circ}$). 
A detailed study of such clouds and complexes is possible based on the constructed map.

Among other things, the map presented here allows the reddening $E(J-K_s)$ to be estimated at each point of the space under consideration through
a trilinear interpolation of the values given in the map. In addition, the constructed $E(J-K_s)$ map can be used to estimate the reddening $E(B-V)$ of specific
stars and the interstellar extinction $A_V$ toward each of them. However, the recalculation of $E(J-K_s)$ to $E(B-V)$ and $A_V$ requires knowing the extinction
law. At present, extensive studies of the extinction law in the space under consideration are being carried out (for a review, see Gontcharov 2016b). To a first
approximation, we can adopt relation (7) in accordance with the PARSEC extinction law and use the 3D map of $R_V$ variations from Gontcharov (2012a)
for $R<600$ pc for the calculations using Eq. (2) by taking $R_V=3.1$ at $R>600$ pc. Thus, in addition to the main map of $E(J-K_s)$ variations, in this
study we also produced the 3D maps of $E(B-V)$ and $A_V$ variations instead of the similarly constructed and now outdated maps from Gontcharov (2012b).
All of the constructed maps are presented in Tables 1 and 2 (published fully in electronic form). In Table 1 the map is present in the following form: one row --
one cubic $20\times20\times20$ pc spatial cell. For each cell it gives the following: the rectangular Galactic coordinates of the cell center $X$, $Y$, and $Z$ in pc,
the cumulative reddening $E(J-K_s)$ toward the cell center, the reddening $E(B-V)$ in the approximation (7), the coefficient $R_V$ interpolated based on the
map from Gontcharov (2012a), and the interstellar extinction calculated from Eq. (2). In Table 2 the map is presented in the following form: one row -- one
spatial cell with a size of 20 pc in $R$, 1$^{\circ}$ in $l$, and 1$^{\circ}$ in $b$. For each cell it gives the following: the spherical Galactic coordinates of the cell center $R$ in pc, $l$, $b$
in degrees and the cumulative reddening $E(J-K_s)$ toward the cell center.

%¹¹¹¹¹¹¹¹¹¹¹¹¹¹¹¹¹¹¹¹¹¹¹¹¹¹¹¹¹¹¹¹¹¹¹¹¹¹¹¹¹¹¹¹¹¹¹¹¹¹¹¹¹¹¹¹¹¹¹¹¹¹¹¹¹¹
\begin{table*}[!h]
\def\baselinestretch{1}\normalsize\normalsize
\caption[]{Map of cumulative reddening $E(J-K_s)$, $E(B-V)$, coefficient $R_V$ , and extinction $A_V$ toward the cell center $XYZ$ (published fully in electronic form)
}
\label{table1}
\[
\begin{tabular}{rrrrrrr}
\hline
\noalign{\smallskip}
$X$ & $Y$ & $Z$ & $E(J-K_s)$ & $E(B-V)$ & $R_V$ & $A_V$ \\
\hline
\noalign{\smallskip}
-1200 & 0 & 0           & 0.237 & 0.392 & 3.10 & 1.22 \\
-1180 & -200 & -80 & 0.168 & 0.278 & 3.10 & 0.86 \\
-1180 & -200 & -60 & 0.181 & 0.300 & 3.10 & 0.93 \\
-1180 & -200 &  -40 & 0.181 & 0.300 & 3.10 & 0.93 \\
-1180 & -200 & -20 & 0.186 & 0.308 & 3.10 & 0.95 \\
\hline
\end{tabular}
\]
\end{table*}

%¹¹¹¹¹¹¹¹¹¹¹¹¹¹¹¹¹¹¹¹¹¹¹¹¹¹¹¹¹¹¹¹¹¹¹¹¹¹¹¹¹¹¹¹¹¹¹¹¹¹¹¹¹¹¹¹¹¹¹¹¹¹¹¹¹

%¹¹¹¹¹¹¹¹¹¹¹¹¹¹¹¹¹¹¹¹¹¹¹¹¹¹¹¹¹¹¹¹¹¹¹¹¹¹¹¹¹¹¹¹¹¹¹¹¹¹¹¹¹¹¹¹¹¹¹¹¹¹¹¹¹¹
\begin{table*}[!h]
\def\baselinestretch{1}\normalsize\normalsize
\caption[]{Cumulative reddening $E(J-K_s)$ map toward the cell center $R$, $l$, $b$ (published fully in electronic form)
}
\label{table2}
\[
\begin{tabular}{rrrr}
\hline
\noalign{\smallskip}
$R$ & $l$ & $b$ & $E(J-K_s)$ \\
\hline
\noalign{\smallskip}
  0 & 0 & 0 & 0.000 \\
20 & 0 & -90 & 0.006 \\
20 & 0 & -89 & 0.006 \\
20 & 0 & -88 & 0.006 \\
20 & 0 & -87 & 0.007 \\
\hline
\end{tabular}
\]
\end{table*}

%¹¹¹¹¹¹¹¹¹¹¹¹¹¹¹¹¹¹¹¹¹¹¹¹¹¹¹¹¹¹¹¹¹¹¹¹¹¹¹¹¹¹¹¹¹¹¹¹¹¹¹¹¹¹¹¹¹¹¹¹¹¹¹¹¹

\section*{Conclusions}

In this study we presented an improved version of the 3D stellar reddening map in a space with a radius of 1200 pc around the Sun and within 600 pc of the
Galactic midplane. As in the previous 2010 and 2012 versions of the map, photometry with an accuracy better than $0.05^m$ in the $J$ and $K_s$ bands for more
than 70 million stars from the 2MASS catalogue is used in the new version. However, the data reduction technique is considerably more complicated. As in
the previous versions, an analysis of the distribution of stars near the MS turnoff on the $(J-K_s)$ -- $K_s$ diagram, where they form a distribution maximum,
provides a basis for the method. The shift of this maximum, i.e., $mode(J-K_s)$, along $(J-K_s)$ and $K_s$, given the spatial variations of the mean de-reddened
color $(J-K_s)_0$ of these stars, is interpreted as a growth of the reddening with increasing distance. As before, the spatial $(J-K_s)$ and $(J-K_s)_0$ variations
are compared with the corresponding results of Monte Carlo simulations. However, in the new version we used refined (in recent years) parameters
of the Galactic model, new theoretical isochrones of stars, and a more realistic extinction law for our simulations. As a result, we found $(J-K_s)_0$ variations
with distance slightly differing from the previous ones. We calibrated the absolute magnitude $M_{K_s}$ as a function of $(J-K_s)_0$ based on data for Gaia
DR1 TGAS stars with the most accurate parallaxes. The main difference between the new and previous methods is that instead of the fixed mean absolute
magnitude $M_{K_s}$, de-reddened color $(J-K_s)_0$, distance $R$, and reddening $E(J-K_s)$ for each cell, the individual values of these quantities for each star are calculated
by iterations when solving the system of equations relating them. This allowed us to increase the random accuracy of the map to $\sigma(E(J-K_s))<0.01^m$ and
its spatial resolution to 20 pc in coordinates $X$, $Y$, and $Z$ and distance $R$ and to 1$^{\circ}$ in longitude $l$ and latitude $b$. The systematic errors of the constructed
map were estimated to be $\sigma(E(J-K_s))\approx0.025^m$, or $\sigma(E(B-V))\approx0.04^m$. Their main sources are the uncertainty in the spatial $(J-K_s)_0$ variations
for the stars under consideration at a $0.02^m$ level and the uncertainty in the extinction law. The reddening estimates based on the constructed map for the spatial
cells under consideration are consistent with the estimates of other authors. The necessity of additional studies of the reddening at high latitudes is obvious in
this case. The constructed map successfully positions the TGAS O--F MS stars on the HR diagram among the theoretical isochrones of the PARSEC database
and, thus, is one of the best maps for the space under consideration. Having analyzed the reddening along each line of sight, we also constructed the differential
reddening (reddening gradient) map within each spatial cell. It will allow the orientation, size, shape, and optical depth of Galactic structures, clouds, and
cloud complexes in the space under consideration to be estimated.

\section*{Acknowledgments}

I thank the referee for the useful remarks. This publication uses data from the Two Micron All Sky Survey (2MASS) and the Hipparcos/Tycho and Gaia
space projects. This study was performed using the Centre de Donn\'ees astronomiques de Strasbourg.
This work has made use of data from the European Space Agency (ESA) mission Gaia (https://www.cosmos.esa.int/gaia), 
processed by the Gaia Data Processing and Analysis Consortium (DPAC, \\
https://www.cosmos.esa.int/web/gaia/dpac/consortium). 
Funding for the DPAC has been provided by national institutions, in particular the institutions participating in the Gaia Multilateral Agreement.

\end{document}